\documentclass[12pt,english,longbibliography,aps,prb,preprint,nofootinbib,superscriptaddress,showpacs,showkeys,12pt,sort&compress]{revtex4}
\usepackage{ae,aecompl}
\usepackage[T1]{fontenc}
\usepackage[latin9]{inputenc}
\usepackage{color}
\usepackage{float}
\usepackage{amsmath}
\usepackage{graphicx}

\makeatletter
\@ifundefined{textcolor}{}
{%
 \definecolor{BLACK}{gray}{0}
 \definecolor{WHITE}{gray}{1}
 \definecolor{red}{rgb}{1,0,0}
 \definecolor{GREEN}{rgb}{0,1,0}
 \definecolor{BLUE}{rgb}{0,0,1}
 \definecolor{CYAN}{cmyk}{1,0,0,0}
 \definecolor{MAGENTA}{cmyk}{0,1,0,0}
 \definecolor{YELLOW}{cmyk}{0,0,1,0}
}

\usepackage{amsfonts}
\usepackage{aecompl}\usepackage{epstopdf}

\date{\today}

\makeatother

\usepackage{babel}
\begin{document}

\title{Temperature fluctuations in canonical systems: Insights from molecular
dynamics simulations}

\author{\noindent J. Hickman}
\email{jhickma3@masonlive.gmu.edu}

\selectlanguage{english}%

\address{Department of Physics and Astronomy, MSN 3F3, George Mason University,
Fairfax, Virginia 22030, USA}

\author{\noindent Y. Mishin}
\email{ymishin@gmu.edu}

\selectlanguage{english}%

\address{Department of Physics and Astronomy, MSN 3F3, George Mason University,
Fairfax, Virginia 22030, USA}
\begin{abstract}
Molecular dynamics simulations of a quasi-harmonic solid are conducted
to elucidate the meaning of temperature fluctuations in canonical
systems and validate a well-known but frequently contested equation
predicting the mean square of such fluctuations. The simulations implement
two virtual and one physical (natural) thermostat and examine the
kinetic, potential and total energy correlation functions in the time
and frequency domains. The results clearly demonstrate the existence
of quasi-equilibrium states in which the system can be characterized
by a well-defined temperature that follows the mentioned fluctuation
equation. The emergence of such states is due to the wide separation
of timescales between thermal relaxation by phonon scattering and
slow energy exchanges with the thermostat. The quasi-equilibrium states
exist between these two timescales when the system behaves as virtually
isolated and equilibrium.
\end{abstract}


\maketitle

\section{Introduction\label{sec:Intro}}

Fluctuations of thermodynamic properties play an important role in
phase transformations and many other physical phenomena and diverse
applications. While fluctuations of energy $E$, volume $V$, number
of particles $N$ and other extensive parameters are well-understood,
controversies remain regarding the nature, or even existence,\citep{Kittel_Thermal_Physics,Kittel:1973aa,Kittel:1988aa}
of fluctuations of intensive parameters such as temperature, pressure
and chemical potentials. In particular, the question of temperature
fluctuations in canonical systems has been the subject of discussions
for over a century (see e.g. van Hemmen and Longtin\citep{van-Hemmen:2013aa}
for a historical overview of the subject).

A number of different views on temperature fluctuations can be found
in the literature, including the following:

(i) Temperature fluctuations in canonical systems is a real physical
phenomenon and can be measured experimentally.\citep{Chiu:1992aa}
If the volume and number of particles in the system are fixed, then\citep{Landau-Lifshitz-Stat-phys,Callen_book_1985,Mishin:2015ab}
\begin{equation}
\left\langle (\Delta T)^{2}\right\rangle =\dfrac{kT_{0}^{2}}{Nc_{v}^{0}},\label{eq:1}
\end{equation}
where $\Delta T=T-T_{0}$ is the deviation of the system temperature
$T$ from the thermostat temperature $T_{0}$, $c_{v}^{0}$ is the
constant-volume specific heat (per particle) at the temperature $T_{0}$,
and $k$ is Boltzmann's constant. The angular brackets $\left\langle ...\right\rangle $
indicate the canonical ensemble average. \textcolor{black}{Assuming
ergodicity, $\left\langle ...\right\rangle $ can be computed by averaging
over a long trajectory in the phase space of the system.}\footnote{\textcolor{black}{By contrast, the temperature $T$ appearing in Eq.(\ref{eq:1})
is defined by averaging over much shorter segments of the trajectory
as discussed later in the paper. }} Spontaneous energy exchanges between the system and the thermostat
bring the system to quasi-equilibrium states in which the temperature
is slightly higher or slightly lower than $T_{0}$. It is also possible
to quantify the cross-correlation between the fluctuating temperature
and the system's total energy by the equation\citep{Landau-Lifshitz-Stat-phys,Callen_book_1985,Mishin:2015ab}

\begin{equation}
\left\langle \Delta E\Delta T\right\rangle =kT_{0}^{2},\label{eq:2}
\end{equation}
where $\Delta E=E-E_{0}$ and $E_{0}$ is the equilibrium energy.

(ii) Temperature of a canonical system is \emph{defined} as the temperature
of the thermostat. Thus, $T\equiv T_{0}$ by definition and the very
notion of temperature fluctuations is meaningless.\citep{Kittel:1973aa,Kittel:1988aa,Kittel_Thermal_Physics}

(iii) While fluctuations of the system energy $E$ are well-defined,
non-equilibrium temperature $T$ is ill-defined.\citep{van-Hemmen:2013aa,Kittel:1988aa}
One can \emph{formally} define $T$ as $T\equiv T_{0}+(E-E_{0})/(Nc_{v}^{0})$,
which makes $T$ just a nominal parameter identical to energy.\citep{van-Hemmen:2013aa}
From this point of view, equation (\ref{eq:1}) contains no new physics
in comparison with the well-established energy fluctuation relation\citep{Landau-Lifshitz-Stat-phys,Callen_book_1985,Mishin:2015ab}
\begin{equation}
\left\langle (\Delta E)^{2}\right\rangle =NkT_{0}^{2}c_{v}^{0}.\label{eq:3}
\end{equation}

(iv) Even for an equilibrium \emph{isolated} system, temperature is
not a well-defined parameter. It can be evaluated by measuring the
system energy and trying to estimate the temperature of the thermostat
with which the system was in equilibrium before being disconnected.\citep{Mandelbrot:1989aa,Falcioni:2011aa}
This reduces the temperature definition to a statistical problem addressed
in the framework of the estimation theory. The statistical uncertainty
associated with the temperature estimate can be interpreted as its
``fluctuation''. 

Recently, thermodynamics-based arguments for the viewpoint (i) have
been put forward as part of a more general thermodynamic fluctuation
theory.\citep{Mishin:2015ab} The goal of the present paper is to
provide additional insights into the nature of temperature fluctuations
by conducting molecular dynamics (MD) simulations of a quasi-harmonic
crystalline solid. As an operational definition, the non-equilibrium
temperature is identified with kinetic energy of the particles averaged
on an appropriate timescale. In Sec.~\ref{sec:Theory} we set the
stage by reviewing the thermodynamic arguments\citep{Landau-Lifshitz-Stat-phys,Callen_book_1985,Mishin:2015ab}
and introducing three timescales of the problem that permit a clear
definition of non-equilibrium temperature. After presenting the simulation
methodology in Sec.~\ref{sec:Methodology}, we report on MD results
for the kinetic, potential and total energy fluctuations and the respective
correlation functions for the solid (Sec.~\ref{sec:Results}). Using
this data, we are able to extract the temperature fluctuations and
verify Eqs.(\ref{eq:1}) and (\ref{eq:2}) \emph{independently} of
Eq.(\ref{eq:3}). In Sec.~\ref{sec:Discussion} we summarize the
results of this work and formulate conclusions.

\section{Theory\label{sec:Theory}}

If a thermodynamic system is disconnected from its environment and
becomes isolated, it reaches thermodynamic equilibrium after a characteristic
relaxation time $\tau_{r}$. For a simple system, the equilibrium
state is fully defined by its energy $E$, volume $V$ and number
of particles $N$. The entropy $S$ of an equilibrium isolated system
is a function of $E$, $V$, and $N$. This function can be established
by equilibrating the isolated system with different values of $E$,
$V$ and $N$ and measuring or computing $S$ for each set of these
parameters. The function $S=S(E,V,N)$ is called the fundamental equation\citep{Willard_Gibbs,Callen_book_1985,Mishin:2015ab}
and incapsulates \emph{all} thermodynamic properties of the substance.
The temperature, pressure and chemical potential are defined by the
fundamental equation as the derivatives $T=1/(\partial S/\partial E)$,
$p=T(\partial S/\partial V)$ and $\mu=-T(\partial S/\partial N)$,
respectively. 

Suppose the isolated system is still in the process of relaxation.
While $E$, $V$ and $N$ are fixed, other thermodynamic properties
can vary. If we mentally partition the system into relatively small
subsystems, their parameters $E$, $V$ and $N$ can vary during the
relaxation. It is important to recognize that the relaxation time
$t_{r}$ of a small subsystem is much shorter than $\tau_{r}$ of
the entire system, at least for short-range interatomic forces. Thus,
there is a certain timescale $t_{q}$ such that 
\begin{equation}
t_{r}\ll t_{q}\ll\tau_{r},\label{eq:4}
\end{equation}
on which the small subsystems remain infinitely close to equilibrium,
even though the entire system is not in full equilibrium. The subsystems
weakly interact with each other across their interfaces, causing a
slow drift of the entire system towards equilibrium. Such virtually
equilibrium subsystems are called \emph{quasi-equilibrium}\citep{Mishin:2015ab}
and the entire isolated system is said to be in a quasi-equilibrium
state.\footnote{Landau and Lifshitz\citep{Landau-Lifshitz-Stat-phys} call the quasi-equilibrium
states ``quasi-stationary'', which may cause some confusion since
the term ``stationary'' is often used to describe steady-state flows
in driven systems.} On the quasi-equilibrium timescale $t_{q}$, the isolated system
can be thought of as equilibrated in the presence of isolating walls
separating its small subsystems. Accordingly, each quasi-equilibrium
subsystem $\alpha$ can be described by a fundamental equation $S_{\alpha}=S_{\alpha}(E_{\alpha},V_{\alpha},N_{\alpha})$,
from which the local temperature, pressure and chemical potential
can be found by $T_{\alpha}=1/(\partial S_{\alpha}/\partial E_{\alpha})$,
$p_{\alpha}=T_{\alpha}(\partial S_{\alpha}/\partial V_{\alpha})$
and $\mu_{\alpha}=-T_{\alpha}(\partial S_{\alpha}/\partial N_{\alpha})$,
respectively. If the number of subsystems is large enough, we can
talk about spatially continuous temperature, pressure and chemical
potential fields. Such fields appear in the standard treatments of
irreversible thermodynamics\citep{De-Groot1984} and are only defined
on the quasi-equilibrium timescale. They evolve during the relaxation
process and eventually become uniform when the entire system reaches
equilibrium. 

Following the fluctuation-dissipation concepts,\citep{Landau-Lifshitz-Stat-phys,Nyquist_1928,Onsager1931a,Onsager1931b,Callen_Welton_1951,Kubo:1966aa,Marconia:2008aa}
one can expect that similar quasi-equilibrium states arise during
equilibrium fluctuations in an isolated system. Accordingly, the fluctuated
states can be described by well-defined local values of the intensive
parameters, including temperature. Again, such local intensive parameters
are only defined on the quasi-equilibrium timescale $t_{q}$.

Turning to canonical fluctuations, consider a small subsystem of an
equilibrium isolated system. Let us call this subsystem a system and
the rest of the isolated system a reservoir. Consider a timescale
$t_{q}$ such that $t_{r}\ll t_{q}\ll\tau_{r}$, where $t_{r}$ is
the relaxation time of the system and $\tau_{r}$ is the global relaxation
time of the system plus reservoir. On this timescale, the system can
be considered as quasi-equilibrium and thus virtually isolated. As
such, it possess all intensive properties mentioned above. Fluctuations
generally occur on all timescales. However, if we monitor the system
properties averaged over the timescale $t_{q}$, then we can talk
about fluctuations of its intensive parameters. In particular, quasi-equilibrium
fluctuations that preserve the system volume and number of particles
(canonical ensemble) include well-defined temperature fluctuations.
As long as the temperature is properly defined on the quasi-equilibrium
timescale, it will satisfy the fluctuation relation (\ref{eq:1}).

We next apply these concepts to a crystalline solid comprising a fixed
number of atoms $N\gg1$. The local relaxation timescale $t_{r}$
can be identified with a typical phonon lifetime. Suppose the solid
is isolated and in equilibrium. Its instantaneous potential energy
$U$ and kinetic energy of the centers of mass of the particles $K$
fluctuate whereas the total energy $E=K+U$ is strictly fixed. The
timescale $t_{K}$ of the kinetic (as well as potential) energy fluctuations
is the inverse of a typical phonon frequency $\bar{f}$: $t_{K}\sim1/\bar{f}$.
Assuming that the solid is nearly harmonic, this timescale is much
shorter than $t_{r}$. The temperature of the solid is fixed at $T=E/3k$
and can be evaluated from the equipartition relation $\left\langle K\right\rangle =3NkT/2$
by monitoring the kinetic energy over a long time $t\gg t_{r}$.

If the same solid is now connected to a thermostat, two types of fluctuation
occur. First, the same fluctuations as in the isolated system, including
the energy exchanges between the phonon modes on the $t_{r}$ timescale.
Second, there will be fluctuations in the total energy of the solid
due to energy exchanges between the solid and the thermostat. The
two types of fluctuation are governed by physically different relaxation
processes: phonon scattering inside the solid in the first case and
heat flow between the solid and the thermostat in the second. The
respective relaxation times, $t_{r}$ and $\tau_{r}$, are significantly
different. Usually $\tau_{r}\gg t_{r}$, i.e., the energy exchanges
with the thermostat occur on a much longer timescale that depends
on the system size, the system/thermostat interface and other factors.
Thus, there is a timescale $t_{q}$ in between, $t_{r}\ll t_{q}\ll\tau_{r}$,
on which the solid remains quasi-equilibrium and can be assigned a
well-defined temperature. We can use the equipartition relation to
find this quasi-equilibrium temperature,
\begin{equation}
T=\dfrac{2\left\langle K\right\rangle _{q}}{3Nk},\label{eq:5}
\end{equation}
where the subscript $q$ indicates that the \textcolor{black}{time}
average must be taken on the quasi-equilibrium timescale $t_{q}$.\textcolor{black}{}\footnote{\textcolor{black}{The reader is reminded that $\left\langle ...\right\rangle $
is the time average over a very long trajectory of the system in the
phase space. By default, the time averaging is performed in the canonical
ensemble (NVT); otherwise the ensemble is indicated as a subscript.
For example, in Section \ref{sec:Results-1} we discuss the time average
$\left\langle ...\right\rangle _{NVE}$ computed in the micro-canonical
(NVE) ensemble. Some observables are averaged over many time intervals
of the same finite length (say, $\theta$). This is indicated in the
subscript, e.g., $\left\langle ...\right\rangle _{\theta}$. $\left\langle ...\right\rangle _{q}$
denotes the time average over a finite time interval on the quasi-equilibrium
timescale $t_{q}$.}} 

If the kinetic energy is averaged over the thermodynamic timescale
$t\gg\tau_{r}$, then the equipartition relation trivially gives the
thermostat temperature
\begin{equation}
T_{0}=\dfrac{2\left\langle K\right\rangle }{3Nk}.\label{eq:6}
\end{equation}
By contrast, the quasi-equilibrium temperature defined by Eq.(\ref{eq:5})
fluctuates around $T_{0}$ and is predicted to satisfy the fluctuation
formula (\ref{eq:1}). We emphasize that Eq.(\ref{eq:5}) defines
$T$ independently of the instantaneous or average values of the total
energy and makes no reference to the specific heat of the substance.\footnote{For example, for a molecular solid the rotational and vibrational
degrees of freedom contribute to $c_{v}^{0}$ but do not appear in
Eq.(\ref{eq:5}), which only includes the kinetic energy of the centers
of mass.} Instead, the temperature fluctuations can be used to \emph{extract}
the specific heat $c_{v}^{0}$. For a classical harmonic solid composed
of atoms (not a molecular crystal), $c_{v}^{0}=3k$ and Eq.(\ref{eq:1})
becomes

\begin{equation}
\left\langle (\Delta T)^{2}\right\rangle =\dfrac{T_{0}^{2}}{3N}.\label{eq:7}
\end{equation}

The key point of this treatment is that the kinetic energy of the
centers of mass of the particles must be averaged over the appropriate
timescale. We caution against using the ``instantaneous temperature''
defined by the instantaneous value of the kinetic energy as $\hat{T}=2K/3Nk$,
as is often done in the MD community. The ``temperature'' $\hat{T}$
so defined essentially represents the kinetic energy $K/N$ itself
up to units. Although this unit conversion can sometimes make the
MD results look more intuitive, it fails to predict the correct temperature
fluctuations. Using the standard canonical distribution, it is easy
to show that for any classical system\citep{Lebowitz1967}
\begin{equation}
\left\langle (\Delta K)^{2}\right\rangle =\dfrac{3N(kT_{0})^{2}}{2},\label{eq:7-1}
\end{equation}
from which
\begin{equation}
\left\langle (\Delta\hat{T})^{2}\right\rangle =2\dfrac{T_{0}^{2}}{3N}.\label{eq:8}
\end{equation}
For an atomic solid, this equation is off by a factor of two. Consequently,
the specific heat of the solid extracted from Eq.(\ref{eq:1}) using
the ``instantaneous temperature'' $\hat{T}$ is $3k/2$ instead of
the correct $3k$. 

In spite of the failure of the ``instantaneous temperature'' $\hat{T}$
to describe the mean-square fluctuation of temperature, it does satisfy
some other fluctuation relations, including Eq.(\ref{eq:2}) which
then becomes $\left\langle \Delta E\Delta\hat{T}\right\rangle =kT_{0}^{2}$.
Like the energy variance $\left\langle (\Delta E)^{2}\right\rangle $,
the covariance $\left\langle \Delta E\Delta T\right\rangle $ remains
the same for both instantaneous and quasi-equilibrium fluctuations.

In the following sections, Eqs.(\ref{eq:1}), (\ref{eq:2}) and (\ref{eq:3})
will be verified by MD simulations with different choices of the thermostat. 

\section{Methodology of simulations\label{sec:Methodology}}

\subsection{Molecular dynamics simulations\label{subsec:MD}}

As a model system we chose face-centered cubic copper with atomic
interactions described by an embedded-atom potential.\citep{Mishin01}
The potential accurately reproduces many physical properties of Cu,
including phonon dispersion relations. The MD simulations were performed
with the LAMMPS code\citep{Plimpton95} with the time integration
step of $dt=0.001$ ps. Except for the system in a ``natural thermostat''
discussed later, all simulations were conducted in a cubic simulation
block with periodic boundary conditions. The block edge was 7.23 nm
and the total number of atoms was $N=32000$. The block edges were
aligned with $\left\langle 100\right\rangle $ directions of the crystal
lattice. The simulation temperature was chosen to be $T_{0}=100$
K and the lattice parameter was adjusted to ensure that the solid
was stress-free at this temperature. 

Prior to studying thermal fluctuations, two types of additional simulations
were performed to generate data needed for a comparison with fluctuation
results. Firstly, the phonon density of states $g(f)$ at 100 K was
computed by the method developed by Kong\citep{Kong2011} and implemented
in LAMMPS. This method was chosen because it does not rely on fluctuations
and provides independent results for comparison. Secondly, to test
the accuracy of the simulation methodology, the specific heat of the
solid was computed by a direct (non-fluctuation) method. This was
accomplished by running canonical (NVT) MD simulations at the temperatures
of 50, 100 and 150 K and calculating the \textcolor{black}{time} average
energies $\left\langle E\right\rangle $. The volume was fixed at
the value corresponding to 100 K. The energy was found to follow a
linear temperature dependence in this temperature interval, from which
the derivative $(\partial\left\langle E\right\rangle /\partial T)_{N,V}$
was evaluated by a linear fit. The specific heat at 100 K was then
found from the equation $c_{v}^{0}=(\partial\left\langle E\right\rangle /\partial T)_{N,V}/N$.
The number obtained was $24.89$ J/(mol~K), which is close to the
equipartition theorem prediction $3k=24.94$ J/(mol~K).

The subsequent MD simulations utilized two ensembles: the microcanonical
NVE (isolated system) and canonical NVT (system in a thermostat).
The NVE system was prepared so that the temperature evaluated from
the relation $\left\langle K\right\rangle _{NVE}=3NkT/2$ was very
close to 100 K. In the NVE ensemble, the MD simulation simply integrates
the classical equations of motion with a Hamiltonian dictated by the
interatomic potential. The NVT simulations utilized the Langevin thermostat
built into LAMMPS.\citep{Plimpton95} The Langevin algorithm\citep{FrenkelS02}
mimics a thermostat by treating the atoms as if they were embedded
in an artificial viscous medium composed of much smaller particles.
This medium exerts a drag force as well as a stochastic noise force
$\boldsymbol{R}$ that constantly perturbs the atoms. The total force
on atom $i$ is 

\begin{equation}
\boldsymbol{F}_{i}=-\frac{\partial U(\boldsymbol{r}_{1}...\boldsymbol{r}_{N})}{\partial\boldsymbol{r}_{i}}-m_{i}\gamma\boldsymbol{v}_{i}+\boldsymbol{R}_{i}.\label{eq:9}
\end{equation}
Here, $U(\boldsymbol{r}_{1}...\boldsymbol{r}_{N})$ is the potential
energy due to atomic interactions, and $m_{i}$, $\boldsymbol{r}_{i}$
and $\boldsymbol{v}_{i}$ are, respectively, the mass, position and
velocity of atoms $i$. The drag term depends on the damping constant
$\gamma$, the inverse of which controls the timescale $\tau_{r}$
of the energy exchanges between the solid and the thermostat. During
the simulation, the noise $\boldsymbol{R}_{i}$ is randomly sampled
from a normal or uniform distribution at time intervals much shorter
than $\tau_{r}=1/\gamma$. The variance of the noise defines the thermostat
temperature $T_{0}$ via the standard fluctuation-dissipation relation.\citep{FrenkelS02}
To evaluate the role of the thermostat, additional simulations were
conducted with a Nose-Hoover thermostat as will be discussed later.

\subsection{Post-processing procedures\label{subsec:Post-processing}}

We next describe the statistical analysis of the MD results at the
post-processing stage. Consider a long MD simulation run implemented
for a time $t_{\mathrm{tot}}$. Suppose two fluctuating properties,
$X$ and $Y$, are saved at every integration step of the simulation.
These can be the kinetic, potential or total energy of the solid.
We trivially compute the \textcolor{black}{time} average values $\left\langle X\right\rangle $
and $\left\langle Y\right\rangle $, as well as the variances $\langle(\Delta X)^{2}\rangle$
and $\langle(\Delta Y)^{2}\rangle$ and the covariance $\langle\Delta X\Delta Y\rangle$,
were $\Delta X=X-\left\langle X\right\rangle $ and $\Delta Y=Y-\left\langle Y\right\rangle $. 

For a spectral analysis, we break the long stochastic processes $X(t)$
and $Y(t)$ into a large number of shorter processes, $x(t)$ and
$y(t)$, by dividing the total time $t_{\mathrm{tot}}$ into smaller
intervals of the same duration $\theta\ll t_{\mathrm{tot}}$. The
time $\theta$ was chosen to be longer than the correlation times
of both variables, so that the intervals represent statistically independent
samples with different initial conditions. For each time interval
$0\leq t\leq\theta$ we perform a discrete Fourier transformation
of $x(t)$ and $y(t)$ to obtain a set of Fourier amplitudes, $\hat{x}_{j}$
and $\hat{y}_{j}$, corresponding to the frequencies $f_{j}=j/\theta$,
where $j=0,\pm1,\pm2,...$. These amplitudes are complex numbers satisfying
the symmetry relations $\hat{x}_{-j}=\hat{x}_{j}^{*}$ and $\hat{y}_{-j}=\hat{y}_{j}^{*}$
(the asterisk denotes complex conjugation). The functions
\[
\hat{C}_{XX}(f_{j})=\frac{\overline{\hat{x}_{j}\hat{x}_{j}^{*}}}{f_{1}},\qquad\hat{C}_{YY}(f_{j})=\frac{\overline{\hat{y}_{j}\hat{y}_{j}^{*}}}{f_{1}},
\]
where the bar denotes averaging over all \textcolor{black}{time} intervals,
represent the ensemble-averaged power spectra of $X$ and $Y$. Likewise,
\[
\hat{C}_{XY}(f_{j})=\frac{\overline{\hat{x}_{j}\hat{y}_{j}^{*}}}{f_{1}}
\]
represents the spectral power of $X$-$Y$ correlations. 

Following the Wiener-Khinchin theorem,\citep{Landau-Lifshitz-Stat-phys,Kubo:1991aa}
the functions $\hat{C}_{XX}(f_{j})$, $\hat{C}_{YY}(f_{j})$ and $\hat{C}_{XY}(f_{j})$
were then subject to inverse Fourier transformations to obtain the
auto-correlation functions (ACF) $C_{XX}(t)=\left\langle X(0)X(t)\right\rangle $
and $C_{YY}(t)=\left\langle Y(0)Y(t)\right\rangle $ and the cross-correlation
function (CCF) $C_{XY}(t)=\left\langle X(0)Y(t)\right\rangle $. In
this work, we are interested in correlations between properties relative
to their average values, namely, $C_{\Delta X\Delta X}(t)=\left\langle \Delta X(0)\Delta X(t)\right\rangle $,
$C_{\Delta Y\Delta Y}(t)=\left\langle \Delta Y(0)\Delta Y(t)\right\rangle $
and $C_{\Delta X\Delta Y}(t)=\left\langle \Delta X(0)\Delta Y(t)\right\rangle $.
These were readily obtained by removing the point $f_{0}$ from the
spectra prior to the Fourier inversion.

All correlation functions in the frequency domain shown in the figures
below have been normalized by $\langle(\Delta X)^{2}(\Delta Y)^{2}\rangle^{1/2}$.
For ACFs, the area under the normalized plots agains the frequency
is therefore unity. 

To evaluate the effect of the averaging timescale on the fluctuation
relations more directly, the spectral analysis was supplemented by
a simple coarse-graining procedure in the time domain. For this procedure,
we lifted the requirement that the time interval $\theta$ be longer
than the correlation time. For every time interval $l$, we computed
the \textcolor{black}{time} average energy values $\left\langle X\right\rangle _{l}$,
$\left\langle Y\right\rangle _{l}$, etc. A formal temperature $T_{l}$
was defined by the equipartition relation $T_{l}=2\left\langle K\right\rangle _{l}/3Nk$.
These coarse-grained values were then treated as a new dataset, for
which we computed the fluctuation properties such as $\langle(\Delta E)^{2}\rangle_{\theta}$,
$\langle(\Delta T)^{2}\rangle_{\theta}$ and $\langle\Delta E\Delta T\rangle_{\theta}$.
These fluctuation properties were examined as functions of the time
interval $\theta$. For $\theta=dt$, this procedure reduces to computing
the fluctuations of instantaneous properties. By increasing $\theta$,
we can scan various timescales, including $t_{r}$, $\tau_{r}$, and
the quasi-equilibrium timescale in between.

\section{Simulation results and discussion\label{sec:Results}}

\subsection{{\normalsize{}NVE simulations\label{sec:Results-1}}}

\begin{figure}[H]
\noindent \begin{centering}
\textbf{(a)}\includegraphics[scale=0.62]{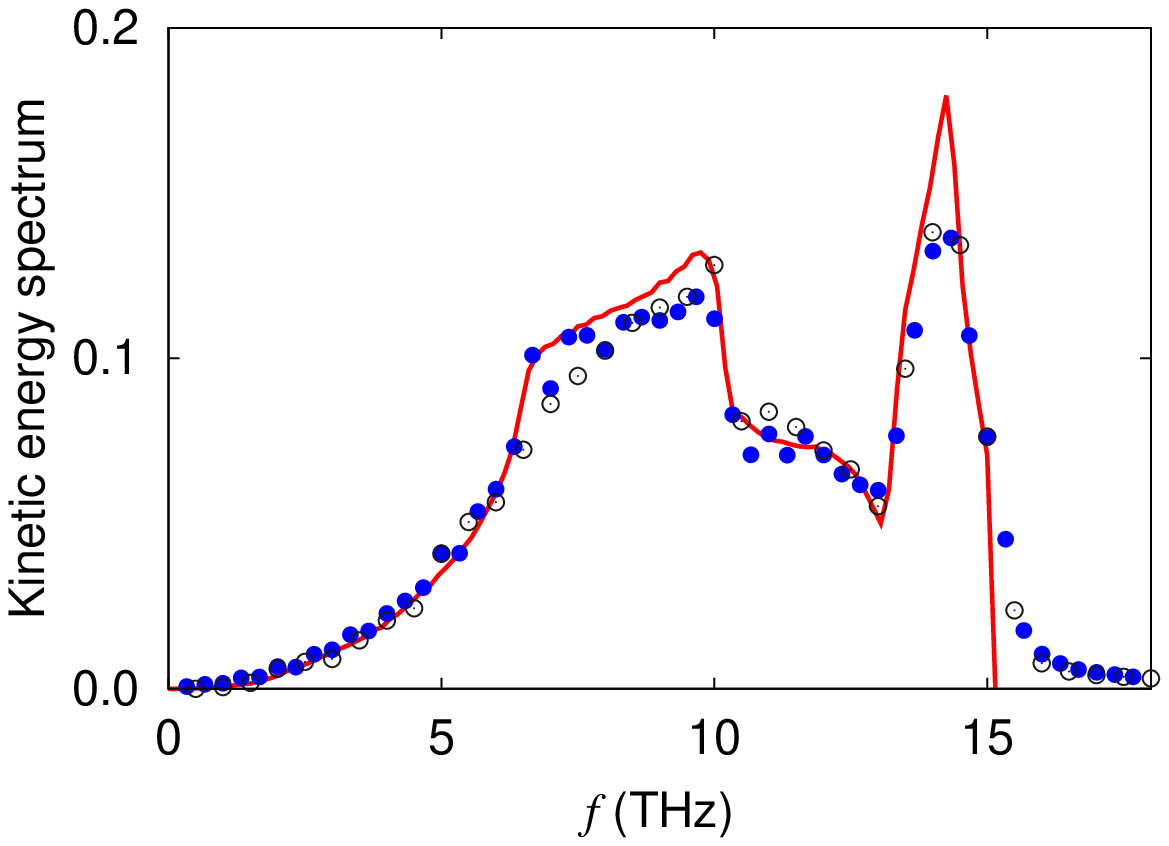}\bigskip{}
\par\end{centering}
\noindent \begin{centering}
\textbf{(b)}\includegraphics[scale=0.62]{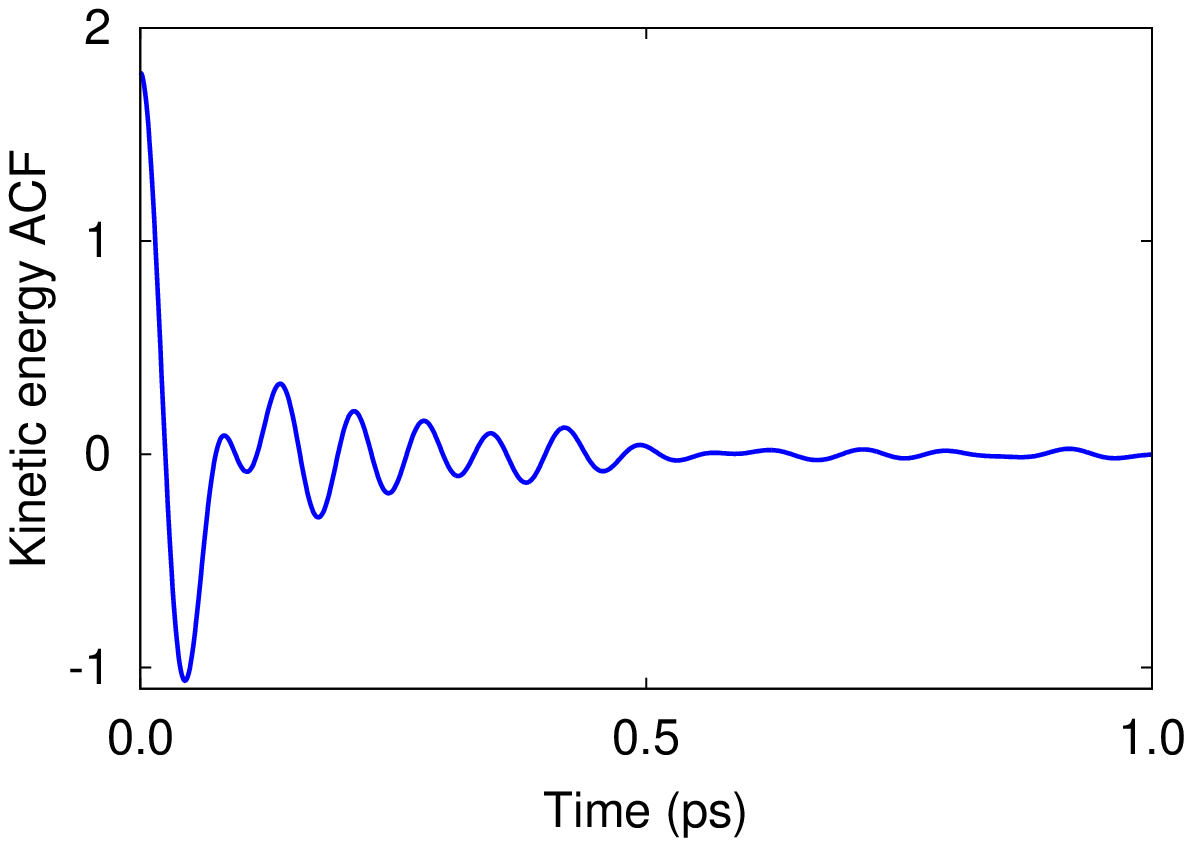}
\par\end{centering}
\caption{Results of NVE MD simulations. (a) Normalized power spectrum $\hat{C}_{\Delta K\Delta K}(f)$
of kinetic energy fluctuations (filled circles), velocity ACF $\hat{C}_{vv}(f/2)$
(open circles), and phonon density of states $g(f/2)$ (solid line).
(b) The kinetic energy ACF $C_{\Delta K\Delta K}(t)$. \label{fig:1}}
\end{figure}

The goal of the NVE simulations was to evaluate the phonon relaxation
time at the chosen temperature and make consistency checks of the
methodology. Figure \ref{fig:1} shows the kinetic energy ACF in the
frequency and time domains. The results were obtained from a $t_{\mathrm{tot}}=2$
ns MD run by averaging over $\theta=3$ ps time intervals. For comparison,
the plot of $\hat{C}_{\Delta K\Delta K}(f)$ {[}Fig.~\ref{fig:1}(a){]}
includes the phonon density of states $g(f/2)$ computed by the non-fluctuation
method\citep{Kong2011} and plotted against the \textcolor{black}{frequency
$f$} followed by normalization to unit area. The close similarity
between the plots is not surprising: in a perfectly harmonic solid,
the kinetic energy ACF is identical to the phonon density of states
except for the doubling of the frequency scale.\citep{Dicky:1969,Scheidler:2001aa,Hickman_to_be_published_2016b}
This doubling is due to the fact that kinetic energy goes through
zero twice per vibration period. In the present simulations, the vibrations
were not perfectly harmonic. The anharmonicity slightly washed out
the shape of the spectrum and produced a high-frequency tail. Since
the total energy is strictly conserved, the potential energy ACF has
an identical shape (not shown here). As another test, the velocity
ACF $\hat{C}_{vv}(f)$ was computed from the same simulation run.
As expected, it was found to be very similar to $\hat{C}_{\Delta K\Delta K}(f)$
except for the frequency doubling effect: $\hat{C}_{vv}(f/2)\approx g(f/2)\approx\hat{C}_{\Delta K\Delta K}(f)$.

The time-dependent ACF $C_{\Delta K\Delta K}(t)$ shown in Fig.~\ref{fig:1}(b)
indicates that the relaxation time due to phonon scattering is about
0.5 ps. Strictly speaking, this time depends on the phonon frequency
and polarization, but we are only interested in a crude estimate.
For comparison, the period $t_{K}$ of kinetic energy fluctuations
can be estimated using a typical frequency of $\bar{f}=10$ THz {[}Fig.~\ref{fig:1}(a){]},
which gives about $t_{K}\approx0.1$ ps. The factor of five difference
between the two timescales is a measure of anharmonicity of this solid
at 100 K.

In the NVE ensemble, the variance of the kinetic energy of the centers
of mass of the particles is\citep{Lebowitz1967} 
\begin{equation}
\langle(\Delta K)^{2}\rangle_{NVE}=\frac{3N(kT_{0})^{2}}{2}\left(1-\frac{3k}{2c_{v}^{0}}\right).\label{eq:13}
\end{equation}
Using $\langle(\Delta K)^{2}\rangle_{NVE}$ obtained by the simulation,
this equation was inverted to solve for $c_{v}^{0}$. The number obtained
was $25.06$ J/(mol~K), which is in good agreement with $24.94$
J/(mol~K) predicted by the equipartition theorem. 

We emphasize that equilibrium temperature fluctuations in the NVE
ensemble are undefined since quasi-equilibrium states are only sampled
by small subsystems of the system but not the system as a whole. As
already mentioned, one can always formally define an ``instantaneous
temperature'' $\hat{T}$ and its fluctuations, but this temperature
is identical (up to units) to the instantaneous kinetic energy per
atom and does not provide new physical insights. 

\subsection{{\normalsize{}NVT simulations\label{sec: Results-2}}}

\begin{figure}[h]
\noindent \begin{centering}
\textbf{(a)}\includegraphics[scale=0.6]{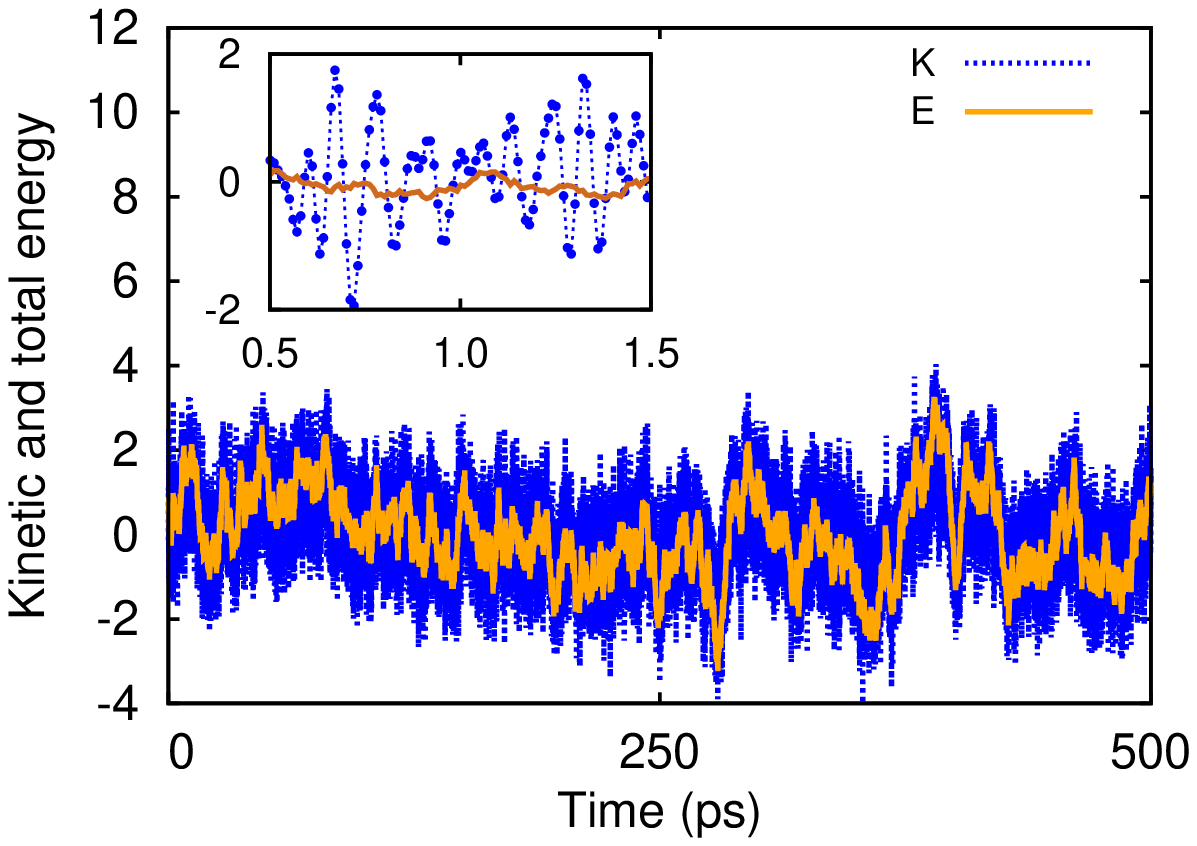}\bigskip{}
\par\end{centering}
\noindent \begin{centering}
\textbf{(b)}\includegraphics[scale=0.6]{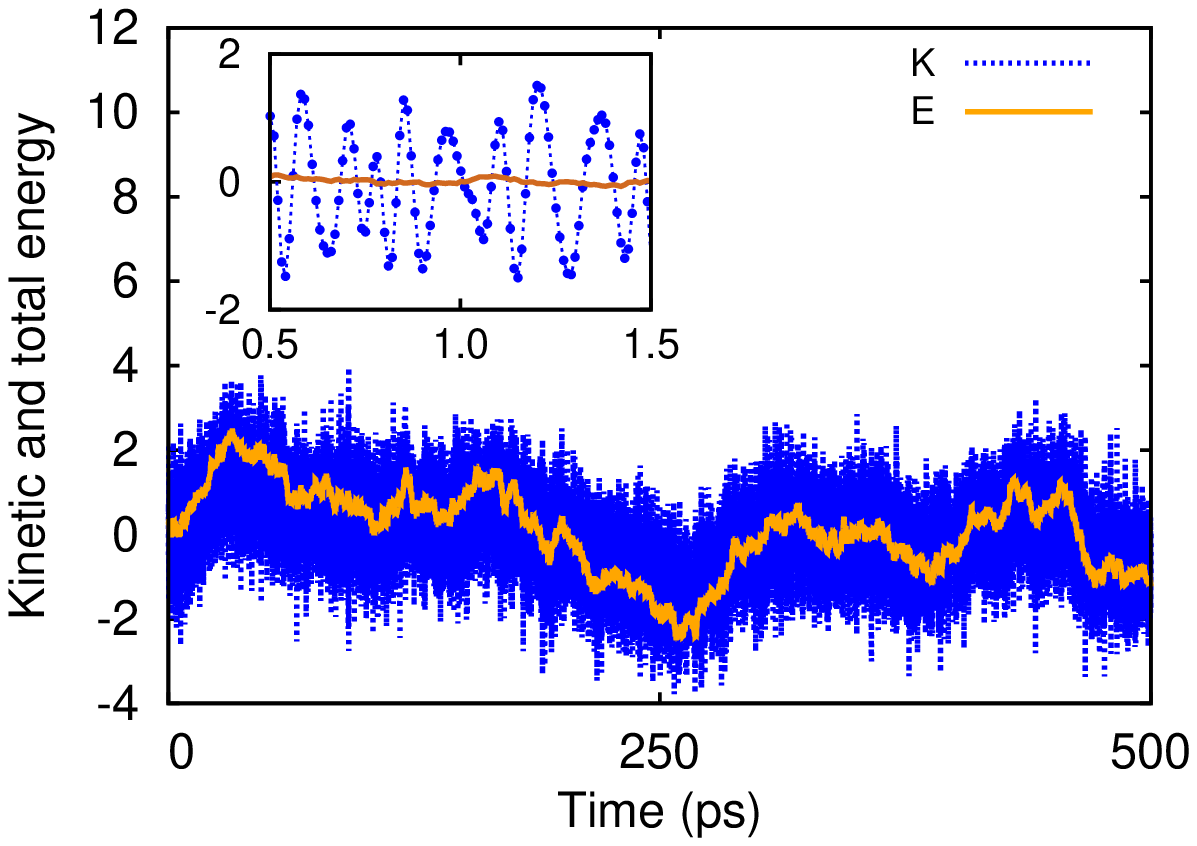}
\par\end{centering}
\caption{Representative fluctuations of the kinetic (blue) and total (orange)
energy in the NVT ensemble with the thermostat time constants (a)
$\tau_{r}=10$ ps and (b) $\tau_{r}=100$ ps. To enable comparison,
the energies were shifted relative to the average values and normalized
by the standard deviations. The insets zoom into shorter time intervals
to demonstrate the existence of two different timescales of the fluctuations
(fast and slow). \label{fig:2} }
\end{figure}

The NVT MD simulations were conducted with two time constants of the
Langevin thermostat: $\tau_{r}=10$ and $100$ ps. The simulations
times were $t_{\mathrm{tot}}=1000\tau_{r}$ (10 and 100 ns, respectively).
The kinetic and total energy fluctuations are illustrated in Fig.~\ref{fig:2}.
To facilitate the comparison, the energies were shifted relative to
their \textcolor{black}{time} average values and normalized by standard
deviations. The plots clearly demonstrate the existence of two different
fluctuation processes: fast fluctuations of kinetic energy and much
slower fluctuations of total energy. The fast fluctuations occur on
the timescale of phonon frequencies, whereas the slow fluctuations
occur on the thermostat timescale $\tau_{r}$. The large disparity
between the two timescales is demonstrated in the insets, where the
kinetic energy fluctuations are superimposed on nearly constant total
energy. This two-scale behavior is especially manifest for the slower
thermostat ($\tau_{r}=100$ ps) and is a clear signature of quasi-equilibrium
states, in which the system behaves as if it were isolated and thus
maintained a constant energy.

\begin{figure}[h]
\noindent \begin{centering}
\textbf{(a)}\includegraphics[scale=0.55]{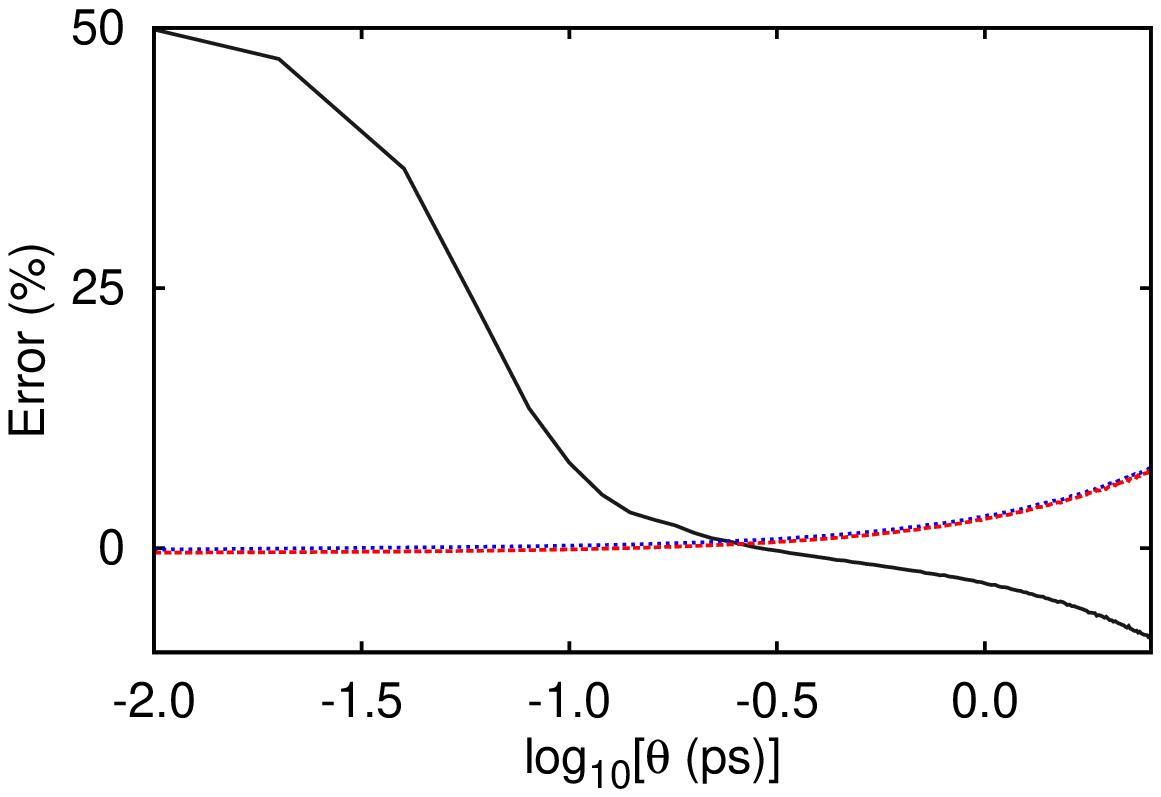}\bigskip{}
\par\end{centering}
\noindent \begin{centering}
\textbf{(b)}~\includegraphics[scale=0.55]{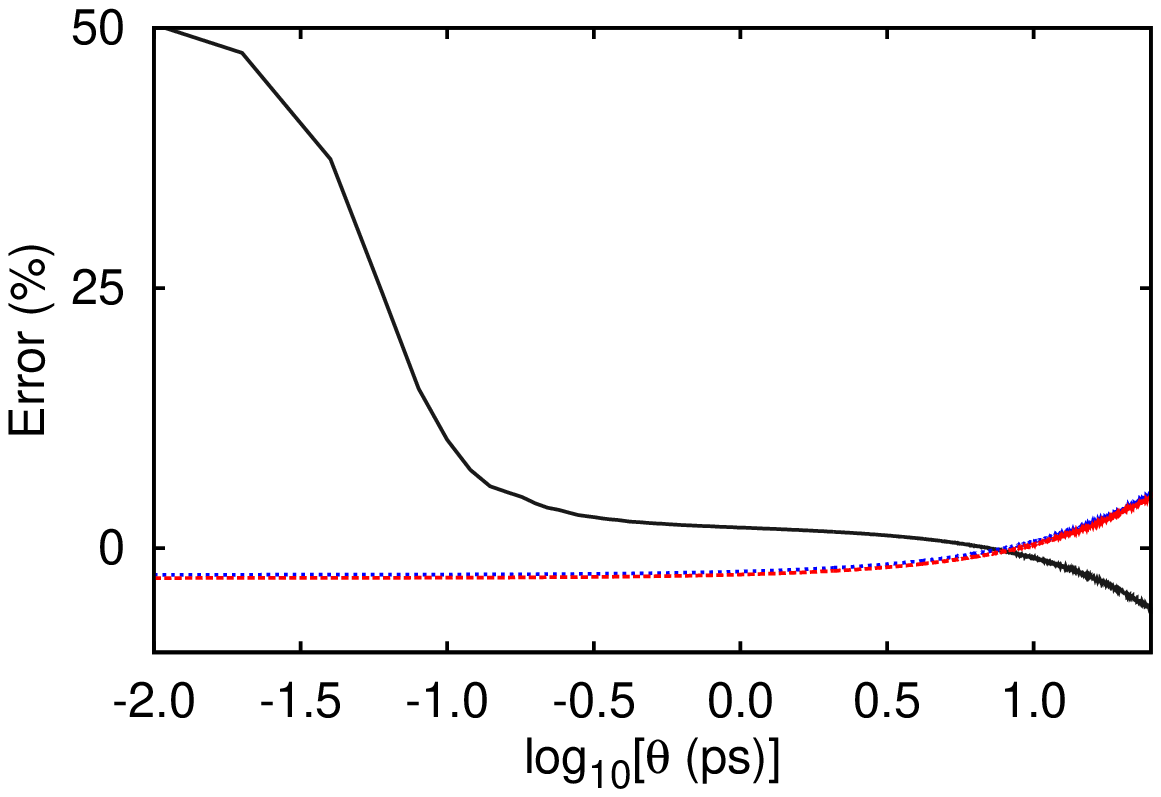}\bigskip{}
\par\end{centering}
\noindent \begin{centering}
\textbf{(c)}\includegraphics[scale=0.55]{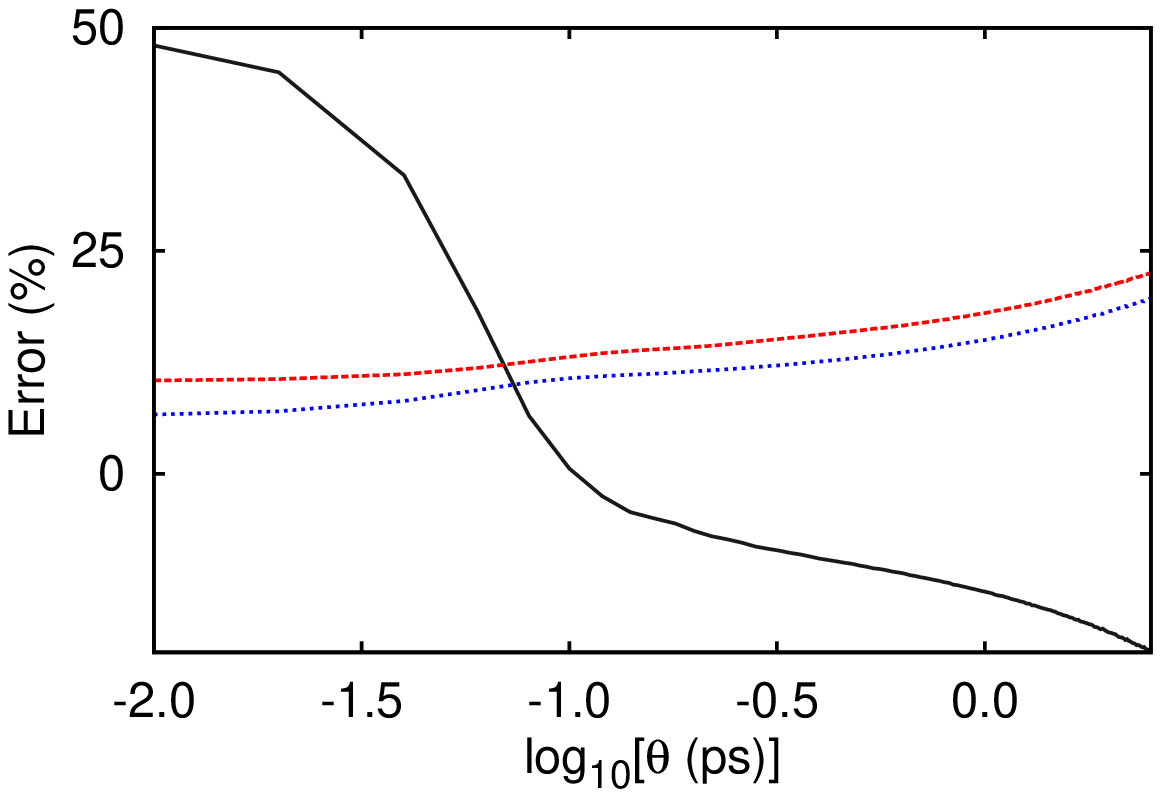}
\par\end{centering}
\caption{Normalized difference between the right and left-hand sides of fluctuation
relations as functions of the averaging time interval $\theta$: Eq.(\ref{eq:1})
(black solid line), Eq.(\ref{eq:2}) (red dashed line) and Eq.(\ref{eq:3})
(blue dotted line) . (a) Langevin thermostat with $t_{r}=10$ ps,
(b) Langevin thermostat with $t_{r}=100$ ps, (c) natural thermostat.
\label{fig:3-1}}
\end{figure}

Figures \ref{fig:3-1}(a,b) show the results of the timescale analysis
discussed in Sec.~\ref{subsec:Post-processing}, in which the energies
were averaged over different time intervals $\theta$ before computing
their fluctuations (Fig.~\ref{fig:3-1}(c) will be discussed later).
The variances/covariances $\langle(\Delta T)^{2}\rangle_{\theta}$,
$\langle\Delta E\Delta T\rangle_{\theta}$ and $\langle(\Delta E)^{2}\rangle_{\theta}$
are compared with the right-hand sides of Eqs.(\ref{eq:1}), (\ref{eq:2})
and (\ref{eq:3}), respectively. The deviation is normalized by the
value of the right-hand side and plotted against $\theta$. Recall
that the minimum value of $\theta$ is the integration step $dt$,
corresponding to instantaneous values of the energies. Observe that
the ``instantaneous temperature'' fluctuation $\left\langle (\Delta\hat{T})^{2}\right\rangle $
has a 50\% error. This number is consistent with the theoretical prediction
in Sec.~\ref{sec:Theory} that an estimate of temperature fluctuations
from $\hat{T}$ will be off by a factor of two. As the averaging time
$\theta$ increases, the error diminishes. When $\theta$ exceeds
the phonon relaxation time $t_{r}$ (about 0.5 ps), the error reduces
to $\pm$ a few percent and remains on this low level until $\theta$
approaches the thermostat time $\tau_{r}$. At that point the error
increases again since the averaging begins to smooth the temperature
fluctuations. In the limit of $\theta\rightarrow\infty$, all fluctuations
are totally suppressed and the error goes to 100\%. This behavior
clearly demonstrates the existence of a timescale on which the temperature
defined by the average kinetic energy satisfies the fluctuation relation
(\ref{eq:1}). As predicted in Sec.~\ref{sec:Theory}, this timescale
lies between $t_{r}$ and $\tau_{r}$ where the system samples quasi-equilibrium
states. Comparing Figs.~\ref{fig:3-1}(a) and \ref{fig:3-1}(b),
we observe that the range of validity of Eq.(\ref{eq:1}) widens as
the thermostat time $\tau_{r}$ increases at a fixed $t_{r}$, which
is again consistent with the definition of quasi-equilibrium states.
By contrast, the errors in $\langle\Delta E\Delta T\rangle_{\theta}$
and $\langle(\Delta E)^{2}\rangle_{\theta}$ remain negligible on
all timescales until $\theta$ approaches $\tau_{r}$ and the averaging
begins to suppress the fluctuations. This is also fully consistent
with the theory. As discussed in Sec.~\ref{sec:Theory}, Eqs.(\ref{eq:2})
and (\ref{eq:3}) remain valid for both instantaneous and quasi-equilibrium
values of the fluctuating properties, which is consistent with Figs.~\ref{fig:3-1}(a,b).

\begin{figure}[h]
\noindent \begin{centering}
\includegraphics[width=0.63\textwidth]{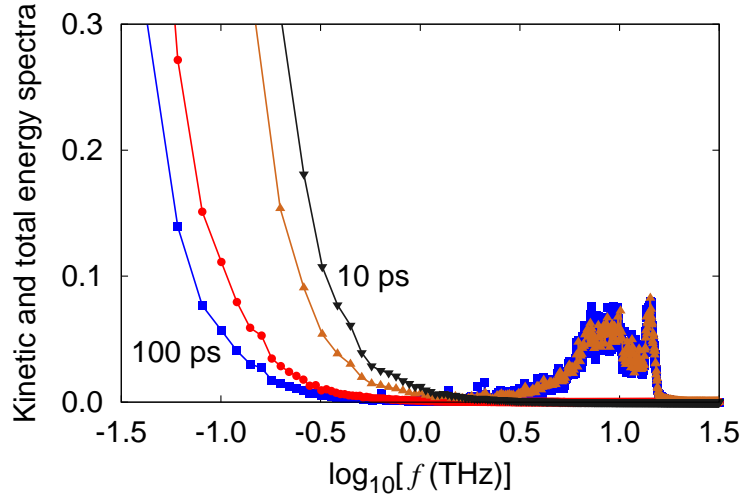}
\par\end{centering}
\caption{Normalized power spectra of kinetic and total energy fluctuations
in the NVT ensemble with a Langevin thermostat for two different time
constants (10 and 100 ps). Square and triangle symbols - kinetic energy,
circle and nabla symbols - total energy. \label{fig:4}}
\end{figure}

Turning to the spectral analysis of the fluctuations, Fig.~\ref{fig:4}
presents the power spectra of the kinetic and potential energies for
the two Langevin thermostats. For the total energy, the spectrum shows
a monotonic decay with frequency and dies off at frequencies larger
than $1/\tau_{r}$, which supports the notion that the total energy
fluctuations are primarily caused by slow exchanges with the thermostat.
By contrast, the kinetic energy spectrum consists of two parts separated
by a frequency gap. The low-frequency part is very similar to that
for the total energy, suggesting a strong correlation. The high-frequency
part has a shape of the phonon spectrum (plotted as a function of
$2f$) and is virtually identical to the spectrum computed in the
NVE ensemble (cf.~Fig.~\ref{fig:1}). Note also that the high-frequency
part of the spectrum is the same regardless of the thermostat time
constant. This part of the spectrum is dominated by the phonon processes
and is independent of how and whether the system interacts with environment.
The gap between the low and high-frequency parts of the spectrum is
where the system is found in quasi-equilibrium states. As expected,
this gap widens as $\tau_{r}$ increases. 

\begin{figure}[h]
\noindent \begin{centering}
\textbf{(a)}\includegraphics[scale=0.68]{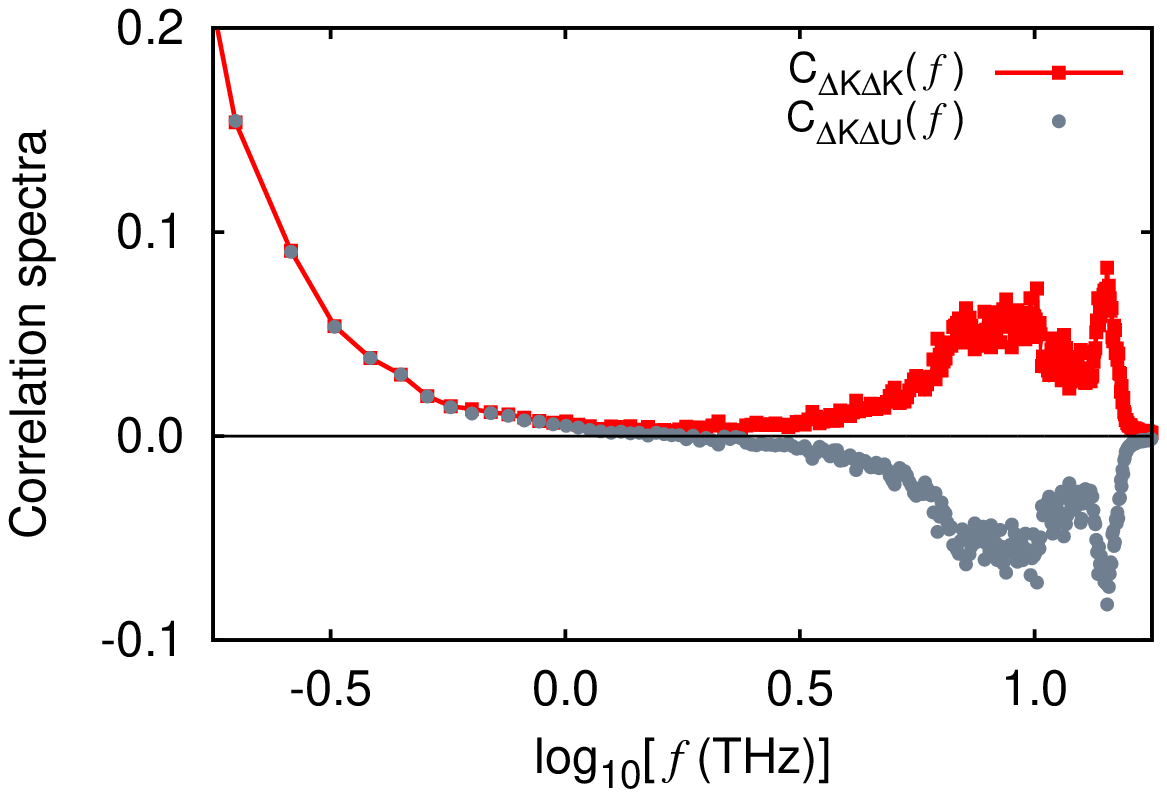}\bigskip{}
\par\end{centering}
\noindent \begin{centering}
\textbf{(b)}\includegraphics[scale=0.68]{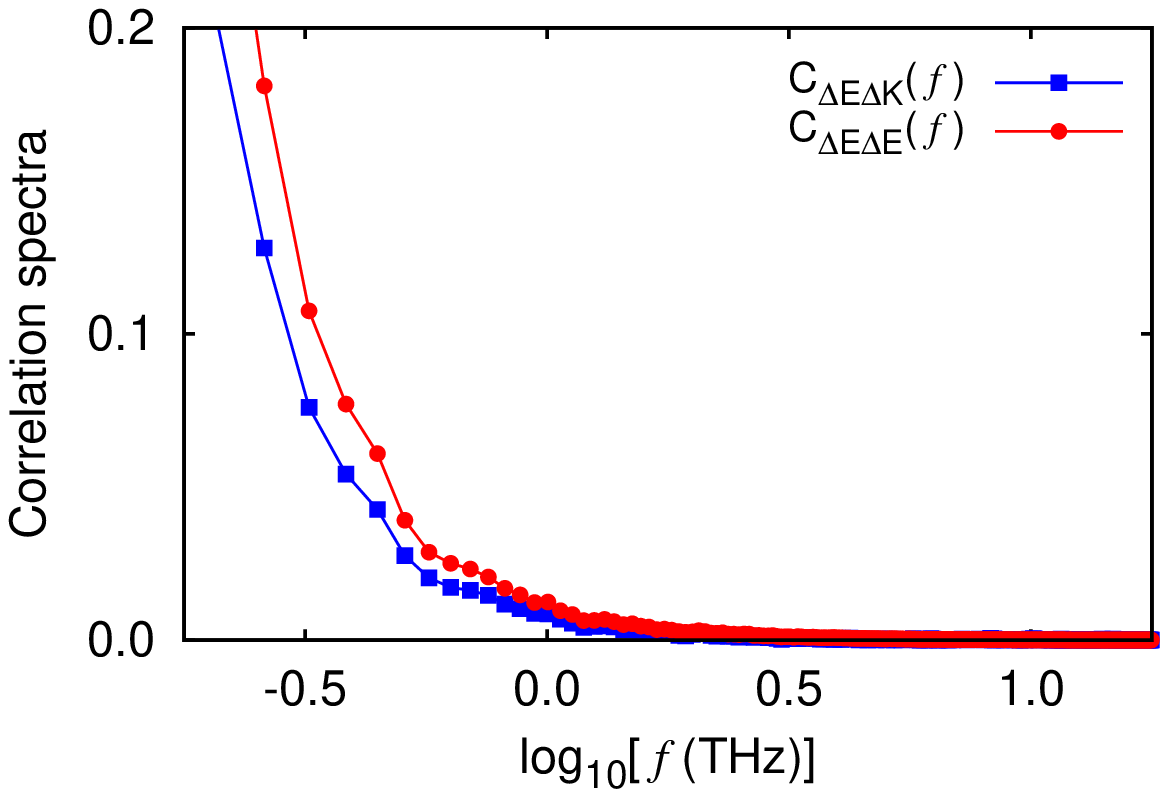}
\par\end{centering}
\caption{Results of NVT MD simulations with a Langevin thermostat ($\tau_{r}=10$
ps). (a) Comparison of the kinetic energy ACF and kinetic-potential
energy CCF in the frequency domain. Note that both spectra have the
same shape but opposite sign at high frequencies and coincide at low
frequencies. (b) Comparison of the total energy ACF and kinetic-total
energy CCF in the frequency domain. Both functions show a similar
monotonic decrease with frequency and die off above $1/\tau_{r}$.
\label{fig:5}}
\end{figure}

The kinetic-potential and kinetic-total CCFs in the frequency domain
are plotted in Fig.~\ref{fig:5}. The respective ACFs are also shown
for comparison. Note that, at high frequencies, the kinetic-potential
energy CCF $\hat{C}_{\Delta K\Delta U}(f)$ is a mirror image of the
kinetic energy ACF $\hat{C}_{\Delta K\Delta K}(f)$ {[}Fig.~\ref{fig:5}(a){]}.
This reflects the nearly perfect anti-correlation between the two
energies on the phonon timescale where the energy exchanges with the
thermostat are negligible and the solid behaves as if it were isolated.
In the low-frequency range below the gap, $\hat{C}_{\Delta K\Delta U}(f)$
and $\hat{C}_{\Delta K\Delta K}(f)$ practically coincide. This is
also expected since the energy exchanges with the thermostat increase
or decrees the kinetic and potential energies (averaged over the phonon
timescale) simultaneously. Although these correlation functions are
only shown for $\tau_{r}=10$ ps, the results for $\tau_{r}=100$
ps look very similar except for a wider frequency gap. On the other
hand, the $\hat{C}_{\Delta K\Delta E}(f)$ and $\hat{C}_{\Delta E\Delta E}(f)$
correlation functions are similar for all frequencies {[}Fig.~\ref{fig:5}(b){]}.
In the low-frequency range, this is consistent with the correlated
behavior of all components of energy during the thermostat exchanges.
At high frequencies, the fast fluctuations of kinetic energy and nearly
constant total energy produce a zero CCF. Since both correlation functions
are strongly dominated by low frequencies, $\langle(\Delta E)^{2}\rangle$,
$\langle\Delta E\Delta K\rangle$ and $\langle\Delta E\Delta T\rangle$
remain the same on both the instantaneous and quasi-equilibrium timescales. 

\begin{figure}[h]
\noindent \begin{centering}
\includegraphics[width=0.57\textwidth]{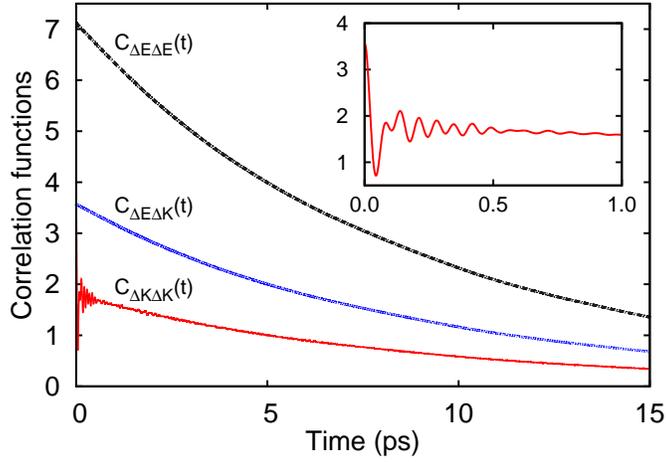}
\par\end{centering}
\caption{Energy correlation functions in the time domain obtained by NVT MD
simulations with a Langevin thermostat ($\tau_{r}=10$ ps). The inset
is a zoom into the short-range part of the kinetic energy ACF. \label{fig:6}}
\end{figure}

\begin{figure}[h]
\noindent \begin{centering}
\includegraphics[width=0.61\textwidth]{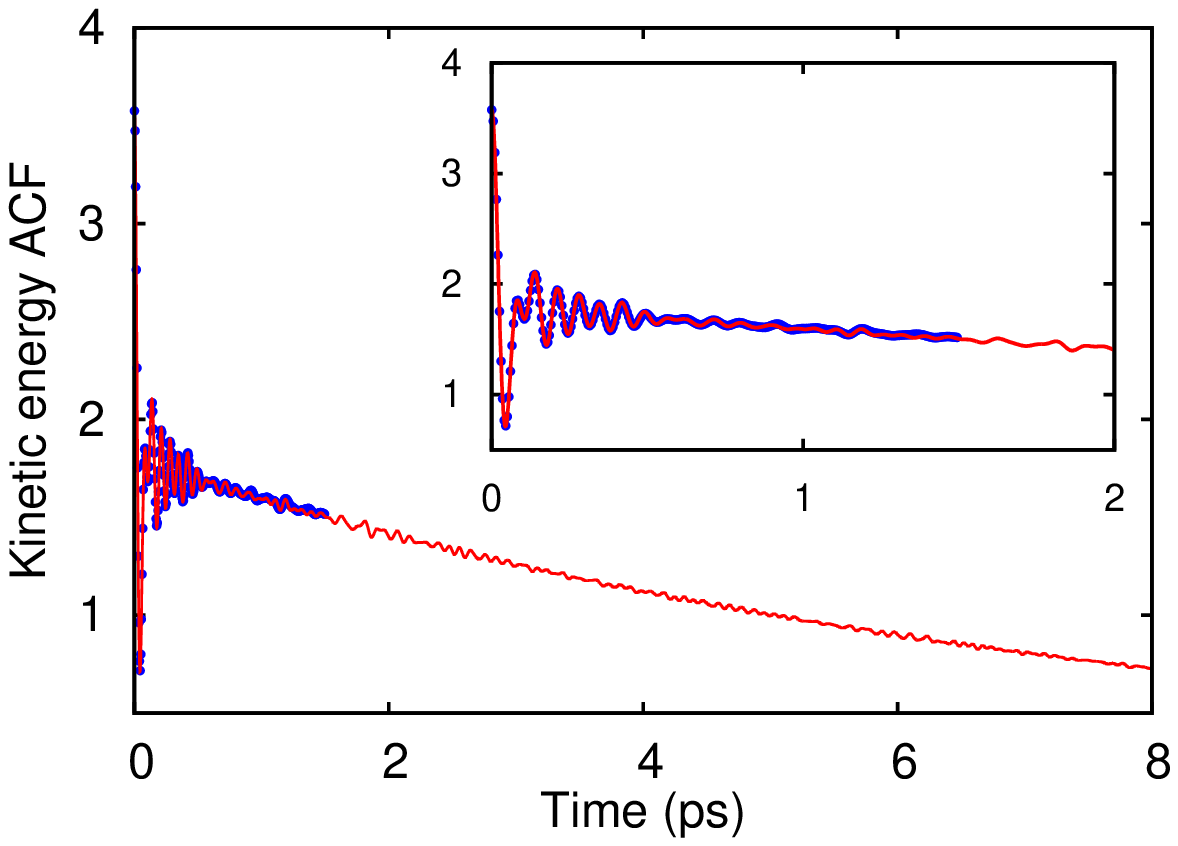}
\par\end{centering}
\caption{The NVT kinetic energy ACF for a Langevin thermostat with $\tau_{r}=10$
ps (red curve) superimposed on the NVE kinetic energy ACF (blue points).
The inset shows a zoom into the short-time region. \label{fig:7-1}}
\end{figure}

Figure \ref{fig:6} shows the correlation functions in the time domain.
Again, only the functions for $\tau_{r}=10$ ps are shown; the result
for $\tau_{r}=100$ ps lead to similar conclusions. Two of the functions
accurately follow the exponential relations
\begin{equation}
C_{\Delta E\Delta E}(t)=\langle(\Delta E)^{2}\rangle e^{-t/\tau_{r}}\label{eq:10}
\end{equation}
and
\begin{equation}
C_{\Delta E\Delta K}(t)=\langle\Delta K\Delta E\rangle e^{-t/\tau_{r}}\label{eq:11}
\end{equation}
expected for a system interacting with a Langevin thermostat. By contrast,
the kinetic energy ACF $C_{\Delta K\Delta K}(t)$ only follows the
exponential relation
\begin{equation}
C_{\Delta K\Delta K}(t)=\langle(\Delta K)^{2}\rangle_{q}e^{-t/\tau_{r}},\enskip t\gg t_{r},\label{eq:15}
\end{equation}
on the timescale $t\gg t_{r}$. Here, $\langle(\Delta K)^{2}\rangle_{q}=1.786$
eV$^{2}$ is the value obtained by extrapolation to $t\rightarrow0$.
For shorter times, $C_{\Delta K\Delta K}(t)$ is a superposition of
Eq.(\ref{eq:15}) and fast-decaying oscillations representing phonon
processes. This short-range part is illustrated in the inset and is
the same for $\tau_{r}=100$ ps (not shown). Furthermore, this part
is identical to $C_{\Delta K\Delta K}(t)$ obtained in the NVE ensemble
(cf.~Fig.~\ref{fig:1}). This is illustrated in Fig.~\ref{fig:7-1}
by superimposing the NVT and NVE ACFs, which show accurate agreement. 

It follows that the entire function $C_{\Delta K\Delta K}(t)$ computed
in the NVT ensemble can be presented in the form
\begin{equation}
C_{\Delta K\Delta K}(t)=\left[C_{\Delta K\Delta K}(t)\right]_{NVE}+\langle(\Delta K)^{2}\rangle_{q}e^{-t/\tau_{r}},\label{eq:17}
\end{equation}
where the first term represents the short-range correlations. Equation
(\ref{eq:17}) shows the same timescale decomposition as already observed
in the spectral form. $\langle(\Delta K)^{2}\rangle_{q}$ represents
the quasi-equilibrium timescale and can be used to calculate the temperature
fluctuations. Taking Eq.(\ref{eq:17}) to the limit of $t\rightarrow0$,
we obtain
\begin{equation}
\langle(\Delta K)^{2}\rangle=\langle(\Delta K)^{2}\rangle_{NVE}+\langle(\Delta K)^{2}\rangle_{q}.\label{eq:18}
\end{equation}
Inserting $\langle(\Delta K)^{2}\rangle$ and $\langle(\Delta K)^{2}\rangle_{NVE}$
from Eqs.(\ref{eq:7-1}) and (\ref{eq:13}), respectively, we arrive
at 
\begin{equation}
\langle(\Delta K)^{2}\rangle_{q}=\frac{9Nk^{3}T_{0}^{2}}{4c_{v}^{0}}.\label{eq:19}
\end{equation}
The temperature is defined by Eq.(\ref{eq:5}), from which
\begin{equation}
\langle(\Delta T)^{2}\rangle=\dfrac{4\langle(\Delta K)^{2}\rangle_{q}}{9N^{2}k^{2}}.\label{eq:20}
\end{equation}
Inserting $\langle(\Delta K)^{2}\rangle_{q}$ from Eq.(\ref{eq:19})
we exactly recover the fluctuation relation (\ref{eq:1}).

As an additional numerical test, $c_{v}^{0}$ was extracted from Eq.(\ref{eq:19})
to obtain $c_{v}^{0}=24.89$ J/(mol~K) in good agreement with the
independent calculation in Sec.~\ref{subsec:MD}. 

\subsection{{\normalsize{}Additional tests\label{subsec:Natural_thermo} }}

To demonstrate that the results reported in the previous sections
are not artifacts of the Langevin thermostat, selected simulations
were repeated using the Nose-Hover thermostat implemented in LAMMPS.\citep{Plimpton95}
The results (not shown here for brevity) were found to be in full
agreement with the simulations employing the Langevin thermostat,
including the timescale separation and validation of the fluctuation
relation (\ref{eq:1}) with temperature computed in quasi-equilibrium
states. 

Both the Langevin and Nose-Hover algorithms implement virtual thermostats
that correctly sample the canonical distribution but still differ
from a physical thermostat. The latter is commonly associated with
a large volume of some inert substance possessing a large heat capacity
and separated from the system by a physical interface. The energy
exchange with the thermostat is then controlled by heat conduction
across the interface, which is different from random perturbations
of atoms uniformly across the system as in the virtual thermostats.
To eliminate any possibility that the virtual thermostats could affect
our conclusions, efforts were taken to model a ``natural'' thermostat
and show that the conclusions remain valid. \textcolor{black}{By a ``natural''
thermostat we mean a simulation block much larger than our system
and separated from the latter by a physical interface.}

As the first step, the NVE MD simulations were executed as above (Sec.~\ref{sec:Results-1}),
but this time, atoms within a relatively small cubic block selected
at the center of the system were treated as the system itself, whereas
the rest of the simulation cell was considered a thermostat. Accordingly,
the energy correlation functions were only computed for the small
subsystem. Repeating the same statistical analyses as above, it was
confirmed that the phonon relaxation time and the thermostat exchange
time were significantly different, creating a large time interval
(accordingly, a frequency gap in the spectrum of kinetic energy) in
which the system existed in quasi-equilibrium states. The temperature
defined on this quasi-equilibrium timescale was found to satisfy the
fluctuation relation (\ref{eq:1}).

But even this test was not found completely satisfactory. The volume
of the inner lattice block selected as our system was not strictly
fixed but rather fluctuated during the simulations. Strictly speaking,
the ensemble implemented on the system was NPT (with zero pressure)
rather than NVT. Although the fluctuation relations (\ref{eq:1})
and (\ref{eq:2}) remain valid in the NPT ensemble as well,\citep{Mishin:2015ab}
the simulations with the virtual thermostats were conducted in a different
(NVT) ensemble.

\begin{figure}
\noindent \begin{centering}
\includegraphics[width=0.7\textwidth]{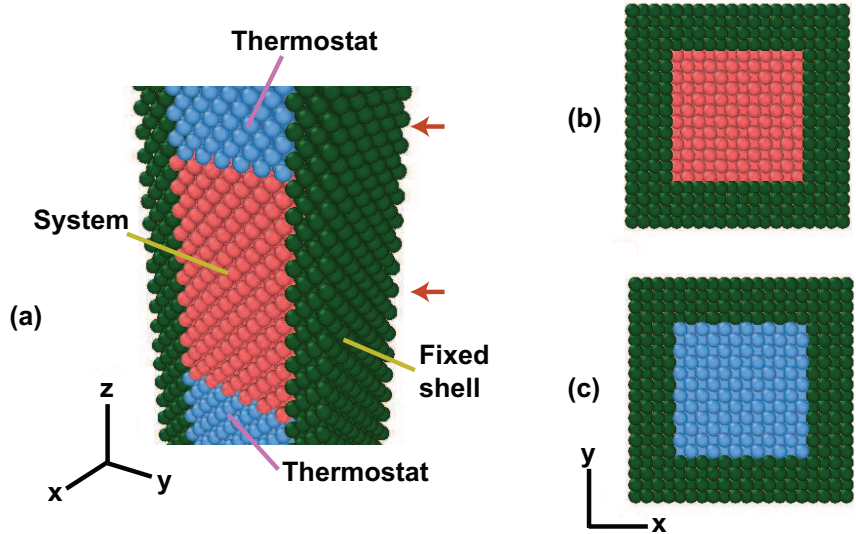}
\par\end{centering}
\caption{The anatomy of the ``natural'' thermostat implemented in this work.
(a) Vertical cross-section of the simulation block revealing the cubic
system under study at the center, the thermostat regions above and
below the system, and a fixed shell enclosing both the system and
the thermostat. The entire assembly is much longer in the vertical
($z$) direction than shown. (b) and (c) show horizontal ($x$-$y$)
cross-sections at the levels indicated by the arrows.\label{fig:8}}
\end{figure}

To make sure that the comparison is made for the same ensemble, the
natural thermostat was redesigned as shown in Fig.~\ref{fig:8}.
A cubic lattice block with an edge of about 2 nm (about 1400 atoms)
was embedded at the center of a larger periodic block with the dimensions
$3.6\times3.6\times72$ nm (80,000 atoms). This relatively small inner
lattice block was the system to be studied. Atoms within a 0.8 nm
shell parallel to the long ($z$) direction were fixed in their positions.
The remaining atoms above and below the cubic block represented the
thermostat and were subject to the following constraint: they could
only vibrate in the $x$ and $y$ directions while their $z$-coordinates
were fixed. As a result, the cubic system was fully surrounded by
atoms incapable of motion in the directions normal to the faces of
the cube. The volume of the system was thereby fixed, imitating rigid
walls of a calorimeter. At the same time, the thermostat atoms above
and below the cube could exchange energy with it by heat conduction
across the interfaces mediated by transverse phonons (polarized in
the $x$-$y$ plane). This heat exchange controlled the system temperature.
The entire assembly was brought to thermal equilibrium at the temperature
of 100 K.\footnote{Since the partially constrained atoms forming the thermostat were
thermally active only in the $x$ and $y$ directions, their temperature
was computed as $\left\langle K\right\rangle /Nk$. In the system
itself, the temperature was as usual $2\left\langle K\right\rangle /3Nk$.} As usual, the lattice parameter was chosen to ensure zero mechanical
stress in the system. Once equilibrium was reached, a 20 ns long NVE
MD simulation was performed to compute statistical properties of fluctuations
as described above. 

Fig.~\ref{fig:3-1}(c) shows the normalized differences between the
variances/covariances $\langle(\Delta T)^{2}\rangle_{\theta}$, $\langle\Delta E\Delta T\rangle_{\theta}$
and $\langle(\Delta E)^{2}\rangle_{\theta}$ computed with the natural
thermostat and the right-hand sides of Eqs.(\ref{eq:1}), (\ref{eq:2})
and (\ref{eq:3}), respectively. The results are qualitatively the
same as obtained with the Langevin thermostat {[}Fig.~\ref{fig:3-1}(a,b){]}.
The deviation from the temperature fluctuation relation (\ref{eq:1})
is again about 50\% when the instantaneous temperature is used ($\theta=dt$)
and reduces to approximately $\pm$ 10\% when the temperature is defined
by the kinetic averaged over the time intervals $\theta\gtrsim0.1$
ps. When $\theta$ reaches a few ps or higher, the error increases
again due to the smoothing of fluctuations by averaging over timescales
comparable with the thermostat time. We can conclude that the latter
must be on the order of 10 ps. Thus, the quasi-equilibrium timescale
for this thermostat is between $\sim0.1$ and $\sim10$ ps. In this
time interval, the temperature fluctuation relation (\ref{eq:1})
is approximately followed, although not as accurately as with the
Langevin thermostat. This is understandable given that the system
in the natural thermostat was a factor of 20 smaller and subject to
a size effect.\footnote{The phonon mean free path at this temperature is estimated to be about
1.3 nm, which is comparable to the system size.} Upscaling of both the system and the thermostat would likely reduce
the error but was not pursued in this work. 

\begin{figure}
\noindent \begin{centering}
\includegraphics[width=0.62\textwidth]{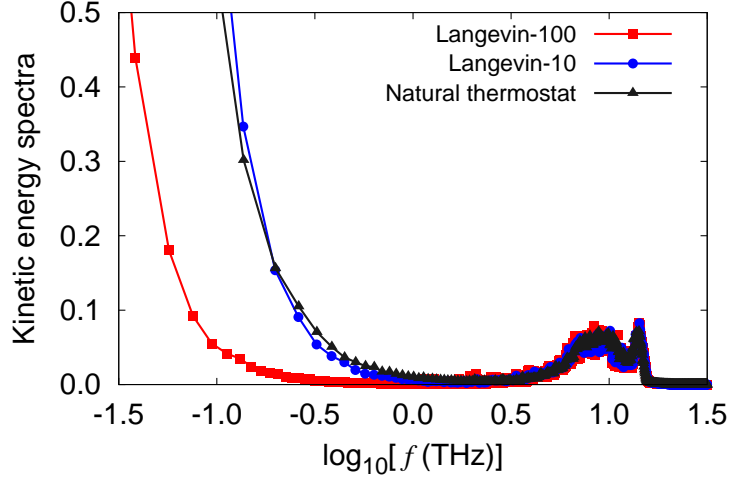}
\par\end{centering}
\caption{Power spectra of kinetic energy from NVT MD simulations of systems
connected to a natural thermostat and two Langevin thermostats with
the time constants of 10 and 100 ps.\label{fig:9} }
\end{figure}

Spectral analysis of energy fluctuations has shown that the system
closely follows the same trends as for the Langevin and Nose-Hoover
thermostats. As one example, Fig.~\ref{fig:9} compares the power
spectra of kinetic energy for the natural and Langevin thermostats.
The high-frequency parts of the spectra coincide almost perfectly.
The low-frequency parts controlled by energy exchanges with the thermostat
also have similar shapes. In fact, for the natural thermostat, this
part of the spectrum is very close to that for the Langevin thermostat
with $\tau_{r}=10$ ps. This confirms the above estimate of the time
constant of the natural thermostat. This also shows that the time
constants of the Langevin thermostat chosen for this study were quite
realistic. Overall, we can conclude that the association of the temperature
fluctuation relation (\ref{eq:1}) with the quasi-equilibrium timescale
has a generic validity and does not reflect some specific features
of thermostats.

\section{Conclusions\label{sec:Discussion}}

We have addressed the long-standing controversy regarding the meaning,
or even existence, of temperature fluctuations in canonical systems.
\textcolor{black}{Over the past decades, the temperature fluctuation
relation (\ref{eq:1}) appearing in many textbooks and papers\citep{Landau-Lifshitz-Stat-phys,Callen_book_1985,Chiu:1992aa,Mishin:2015ab,Hickman_to_be_published_2016b}
has received different interpretations, including the assertion that
this equation is meaningless\citep{Kittel_Thermal_Physics,Kittel:1973aa,Kittel:1988aa}
or at best a mere formality.\citep{Mandelbrot:1989aa,Falcioni:2011aa,van-Hemmen:2013aa}
We have demonstrated that Eq.(\ref{eq:1}) is a physically meaningful
relation that remains valid as long as the temperature is defined
on an appropriate timescale. This interpretation of temperature fluctuations
has been supported by MD simulations of a quasi-harmonic solid connected
to a thermostat. }

The simulations have confirmed the existence of two different fluctuation
timescales in canonical systems. The shorter timescale is associated
with the time required for a small isolated system to reach thermodynamic
equilibrium. For an atomic solid studied here, this time $t_{r}$
is controlled by phonon scattering. In this work, this time was about
0.5 ps at the temperature of 100 K. The longer timescale arises due
to slow energy exchanges between the system and the thermostat. Such
exchanges may occur by a variety of physically different mechanisms,
such as heat transfer across the system/thermostat interface. For
the natural and virtual thermostats studied here, the energy exchange
time $\tau_{r}$ was on the order of 10 to 100 ps. Thus, $\tau_{r}$
is orders of magnitude longer than $t_{r}$. At the intermediate timescale
$t_{q}$ ($t_{r}\ll t_{q}\ll\tau_{r}$) the system remains in internal
thermodynamic equilibrium and can be treated as if it were disconnected
from the thermostat. In such quasi-equilibrium states, it has well-defined
intensive properties such as temperature, pressure and chemical potential. 

In particular, temperature can be defined through the equipartition
relation using the kinetic energy averaged on the quasi-equilibrium
timescale $t_{q}$. It has been shown that fluctuations of the temperature
so defined do follow Eq.(\ref{eq:1}). Attempts to define temperature
through kinetic energy averaged over shorter ($<t_{r}$) or longer
($>\tau_{r}$) time intervals result in significant deviations from
Eq.(\ref{eq:1}). \textcolor{black}{In particular, the ``temperature''
obtained by averaging the kinetic energy over a long time $t\gg\tau_{r}$
does not fluctuate and approaches the thermostat temperature $T$$_{0}$.}

The timescale separation is also reflected in the shape of the kinetic
energy ACF in the frequency domain, showing two peaks separated by
a frequency gap. The peak at $f=0$ arises from energy exchanges with
the thermostat, whereas the second peak is associated with phonon
processes and has the shape of the phonon density of states (plotted
against $2f$). The frequency gap represents the quasi-equilibrium
states. The potential energy ACF has a similar structure and can also
be used for the identification of quasi-equilibrium states. Thus,
measured or computed energy spectra of a canonical system carry all
information about the timescale on which temperature fluctuations
are well-defined and follow Eq.(\ref{eq:1}).

The conclusions of this work were tested by MD simulations with two
virtual thermostats (Langevin and Nose-Hoover) and a natural thermostat
consisting of large crystalline regions surrounding the system. In
the future, a similar study could evaluate the validity of pressure
fluctuation relations for canonical systems.\citep{Landau-Lifshitz-Stat-phys,Rudoi:2000aa,Mishin:2015ab} 

\bigskip{}

\emph{Acknowledgments -} This work was supported by the U.S. Department
of Energy, Office of Basic Energy Sciences, Division of Materials
Sciences and Engineering, the Physical Behavior of Materials Program,
through Grant No.~DE-FG02-01ER45871.


\end{document}